\documentclass{article}
\usepackage{graphicx}
\usepackage{amsfonts}
\usepackage{amssymb}
\usepackage{latexsym}
\newcommand\R{\mathbb{R}}
\newcommand\Sph{\mathbb{S}}
\newenvironment{pf}{\unskip{\bf Proof:}}{\unskip{\hfill $\Box$}}
\def\noproof{\unskip{\hfill $\Box$}}
\newcommand{\lemlab}[1]{\label{lemma:#1}}
\newcommand{\theolab}[1]{\label{theo:#1}}

\newcommand{\eqlab}[1]{\label{eq:#1}}

\newcommand{\figlab}[1]{\label{fig:#1}}
\newcommand{\seclab}[1]{\label{section:#1}}

\newcommand{\lemref}[1]{\ref{lemma:#1}}

\newcommand{\theoref}[1]{\ref{theo:#1}}

\newcommand{\figref}[1]{\ref{fig:#1}}
\newcommand{\eqref}[1]{(\ref{eq:#1})}
\newcommand{\secref}[1]{\ref{section:#1}}

\newtheorem{theorem}{Theorem}
\newtheorem{lemma}{Lemma}

{\catcode`\@=11
\gdef\setft#1#2#3{%
\def\@oddfoot{
{\setbox0=\hbox{#1}
\setbox1=\hbox{#3}
\ifdim\wd0>\wd1
\dimen0=\wd0
\box0\hfil#2\hfil\hbox to\dimen0{\hfil\hfil\box1}
\else \dimen0=\wd1
\hbox to\dimen0{\box0\hfil }\hfil#2\hfil\box1 \fi
}}} }

\def\complaint#1{}
\def\withcomplaints{
\newcounter{mycomplaints}
\def\complaint##1{\refstepcounter{mycomplaints}%
\ifhmode%
\unskip%
{\dimen1=\baselineskip \divide\dimen1 by 2 %
\raise\dimen1\llap{\tiny -\themycomplaints-}}\fi%
\marginpar{\tiny [\themycomplaints]: ##1}}%
}

\let\oldendpf=\endpf
\def\endpf{\oldendpf\par\medskip}

\def\C{{\cal{C}}}
\def\F{{\cal{F}}}
\def\a{{\alpha}}
\def\b{{\beta}}

\def\d{{\delta}}
\def\t{{\theta}}
\def\tri{{\triangle}}
\def\B0{{{\rm Ob}(v_0)}} 
\def\D1{{{\rm Ob}(v_1)}} 

\title{\bf Polygonal Chains Cannot Lock in 4D}
\author{%
Roxana Cocan and Joseph~O'Rourke\thanks{
Dept. of Computer Science, Smith Col\-lege, North\-ampton, 
MA 01063, USA.
\{rcocan,\-orourke\}@cs.\-smith\-.edu.
Research supported by NSF Grant CCR-9731804.
Results first reported in~[CO99]. 
}
}
\begin{document}
\maketitle
\begin{abstract}
We prove that, in all dimensions $d \ge 4$, every simple open polygonal
chain and every tree may be straightened, 
and every simple closed polygonal chain
may be convexified.  These reconfigurations
can be achieved by algorithms that
use polynomial time in the number of vertices,
and result in a polynomial number of ``moves.''
These results contrast to those known for $d=2$,
where trees can ``lock,'' and for $d=3$, where open and
closed chains can lock.
\end{abstract}

\vspace{5cm}
\begin{center}
{\bf Smith Technical Report 063}\\
(Major revision of the August 1999 version with the same report number.)\\
\end{center}

\newpage
\tableofcontents
\newpage
\pagenumbering{arabic}
\setcounter{page}{1}

\section{Introduction}
\subsection{Summary}
A {\em polygonal chain\/} $P=(v_0,v_1,\ldots,v_n)$ is a sequence
of consecutively joined segments
$s_i =v_iv_{i+1}$ of fixed lengths
$\ell_i = |s_i|$,
embedded in space.
A chain is {\em closed\/} if
the line segments are joined in cyclic fashion, i.e., if $v_n=v_0$;
otherwise, it is {\em open}.
A {\em polygonal tree\/} is a collection of segments joined into
a tree structure.
A chain or tree is {\em simple\/} if only adjacent edges intersect,
and only then at the endpoint they share.
We study reconfigurations of simple polygonal chains and trees,
continuous motions that preserve the lengths of all edges
while maintaining simplicity.
One basic goal is to determine if an open chain can be
{\em straightened\/}---stretched out in a straight line,
and whether a closed chain can be {\em convexified\/}---reconfigured
to a planar convex polygon.
For trees, straightening permits noncrossing violations of
simplicity to allow the segments to align along the common straight line.
If an open chain or tree cannot be straightened, or
a closed chain convexified, it
is called {\em locked}.
This terminology is borrowed from~\cite{bddlloorstw-lupc3d-99}
and~\cite{bddloorsw-orfltcl-98}.%
\footnote{
        Straightening for trees is never defined
        in~\cite{bddloorsw-orfltcl-98}.
        Instead they rely on mutually unreachable simple configurations.
}

Most of the work in this area was fueled by the longstanding
open problem of determining whether
every open (or closed) chain
in 2D can be straightened (or convexified).
This was recently settled~\cite{cdr-epcbu-00} in the affirmative:
2D chains cannot lock.
In contrast it was earlier established that trees in 
2D~\cite{bddloorsw-orfltcl-98},
and 
both open and closed chains in 3D~\cite{cj-nepiu-98,bddlloorstw-lupc3d-99}
can lock.
In this paper we prove that, for all dimensions $d \ge 4$,
neither chains (open or closed) nor trees can lock.
We partition our results into four main theorems:

\begin{theorem}
Every simple open chain in 4D may be straightened,
by an algorithm that runs in $O(n^2)$ time
and $O(n)$ space,
and which accomplishes the straightening in $O(n)$ moves.
\theolab{open.4D}
\end{theorem}
Here ``move'' is used in the sense defined in~\cite{bddlloorstw-lupc3d-99}.\footnote{
	``During each move,
	a (small) constant number of individual joint moves occur,
	where for each
	a vertex $v_{i+1}$ rotates monotonically
	about an axis through joint $v_i$,
	with the axis of rotation
	fixed in a reference frame attached to some edges.''
}
Essentially each move is a simple monotonic rotation of a few joints.
We have implemented this algorithm for the case when the
vertices are in general position, when it is straightforward.

Nearly the same algorithm proves the same result for trees,
within the same bounds:
\begin{theorem}
Every simple tree in 4D may be straightened,
by an algorithm that runs in $O(n^2)$ time
and $O(n)$ space,
and which accomplishes the straightening in $O(n)$ moves.
\theolab{tree.4D}
\end{theorem}

\noindent
Closed chains require more effort:
\begin{theorem}
Every simple closed chain in 4D may be convexified,
by an algorithm that runs in $O(n^6 \log n)$ time,
and which accomplishes the straightening in $O(n^6)$ moves.
\theolab{closed.4D}
\end{theorem}

\noindent
All these results easily extend to higher dimensions:
\begin{theorem}
Theorems~\theoref{open.4D}, \theoref{tree.4D},
and~\theoref{closed.4D}
hold for all dimensions $d \ge 4$, i.e., neither polygonal
chains nor trees can lock in dimensions greater than three.
\theolab{d.ge.4}
\end{theorem}


We summarize our results in the context of earlier work in 
the table below.
\begin{center}\begin{tabular}{| c | c | c | }
	\hline
Dimension
	& Chains & Trees
	\\ \hline \hline
$2$	&	Cannot lock     & Lockable
	\\ \hline
$3$	&	Lockable & Lockable
	\\ \hline
$d \ge 4$	&	{\em Cannot lock\/} & {\em Cannot lock\/}
	\\ \hline
\end{tabular}
\end{center}

\subsection{Background}
\seclab{Background}
Before commencing with our technical arguments, 
we start with some background,
with the intent of providing
intuition to support our results.

\paragraph{No Knots in 4D.}
In~\cite{cj-nepiu-98}
and~\cite{bddlloorstw-lupc3d-99}, the same example of a locked
open chain in 3D is provided.
The version in the latter paper is shown in Fig.~\figref{knitting}.

\begin{figure}[htbp]
\centering
\includegraphics[width=10cm]{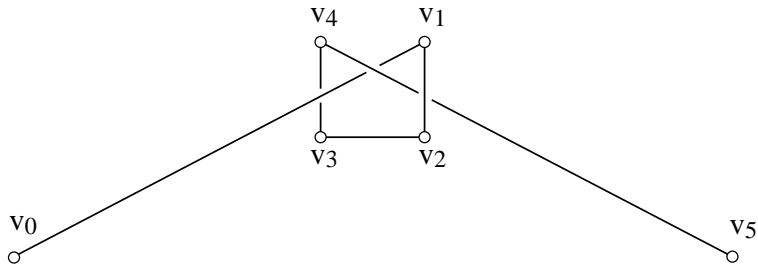}
\caption{The ``knitting needles'' example, based on Fig.~1 in
\protect\cite{bddlloorstw-lupc3d-99} (by permission).
}
\figlab{knitting}
\end{figure}
One proof (used in~\cite{bddlloorstw-lupc3d-99}) that this chain $K$
is locked depends on closing the chain by connecting $v_0$ to $v_5$
to form $K'$,
and then arguing that $K$ can be straightened
iff the corresponding trefoil knot $K'$ can be unknotted, which of course
it cannot.
Thus there is a close connection in 3D between unknotted, locked chains and
knots.
However, the following theorem is well known:
\begin{theorem}
No 1D closed, tame,\footnote{
        A curve is {\em tame\/} if it is topologically
        equivalent to a polygonal curve~\cite[p.5]{cf-ikt-65}.
        Any curve that is continuously differentiable,
        i.e., in class $C^1$, is tame.
}
non-self-intersecting curve $C$
is knotted in $\R^4$.
\end{theorem}

See, e.g.,~\cite[pp.270-1]{a-kb-94}
for an informal proof.
Because proofs of this theorem employ topological deformations,
it seems they are not easily modified to help
settle our questions about chains in 4D.
The rigidity of the links prevents any easy translation of
the knot proof technique to polygonal chains.
However, it does suggest that it would be difficult to
construct a locked chain by extending the methods used in 3D.

\paragraph{No Cages in 4D.}
A second consideration lends support to the intuition
behind our main claim.  This is the inability to confine
one segment in a ``cage'' composed of other segments in 4D.
Consider segment $s_0 = v_0 v_1$ in Fig.~\figref{knitting}.
It is surrounded by other segments in the sense that
it cannot be rotated freely about one endpoint
(say $v_0$) without colliding with the other segments.
Let $S$ be the $2$-sphere in $\R^3$
of radius $\ell_0$ centered at $v_0$.
Each point on $S$ is a possible location for $v_1$.
Segment $s_0$ is confined in the sense that there are points
of $S$ that cannot be reached from $s_0$'s initial position
without collision with the other segments.
This can be seen by centrally projecting the segments from $v_0$ onto $S$,
producing an ``obstruction diagram.''
It should be clear that $v_1$ is confined to a cell of this diagram.
Although this by no means implies that the chain in Fig.~\figref{knitting}
is locked, it is at least part of the reason that the chain might
be locked.

We now argue informally that such confinement is not possible in 4D.
Again let $s_0 = v_0 v_1$ be fixed at $v_0$, and let $S$ be the
$3$-sphere in $\R^4$ of radius $\ell_0$ centered on $v_0$
that represents the possible locations for $v_1$.
Again we project the other segments onto $S$ producing an obstruction
diagram.  As in the lower dimensional case, this diagram is
composed of 1D curves, being the projection of 1D segments.
But in the $3$-sphere $S$, $v_1$ has three degrees of freedom,
and cannot be confined by a
(finite) set of 1D curves.
Our next task is to make this intuitive argument more precise.

\section{Straightening Open Chains in 4D}
\seclab{Open}
Let $P$ be a simple, open polygonal chain in 4D with
$n \ge 2$ vertices.
Each vertex $v_i$ is also called a {\em joint\/} of the chain.
The segment $s_i = v_i v_{i+1}$ we sometimes call a {\em link\/} of the chain.
We say a joint $v_i$ is {\em straightened\/} if
$(v_{i-1}, v_i, v_{i+1})$ are collinear and 
form a simple chain;
in this case, the angle at $v_i$ is $\pi$.

We prove Theorem~\theoref{open.4D} by straightening the first
joint $v_1$, ``freezing'' it, and repeating the process until
the entire chain has been straightened.  
This is a procedure which, of course,
could not be carried out in 3D.  But there is much more room for
maneuvering in 4D.
We have two different algorithms for accomplishing this task.
The first (Algorithm~1a) is easier to understand, but 
only establishes a bound of $O(n^4)$ on the number of moves,
and requires $O(n^4 \log n)$ time. 
The second (Algorithm~1b) is a bit more
intricate but achieves $O(n)$ moves
in $O(n^2)$ time.  
Both follow the rough outline just sketched.
We provide full details for Algorithm~1a, but only sketch Algorithm~1b.

Define the {\em goal position\/} $v_g$ for $v_0$
(and $s_g=v_g v_1$ the goal position for $s_0$)
as the unique
position that represents straightening of joint $v_1$.
Call the goal position {\em intersected\/} if
$s_g \cap s_i \neq \emptyset$ for some $i > 2$;
and otherwise call it {\em free}.

\subsection{Algorithm 1a}
A high-level view of the algorithm is as follows:

\begin{center}
\fbox{%
\begin{minipage}{56mm}
\begin{tabbing}
\hspace*{3mm}\=\hspace*{3mm}\=\hspace*{3mm}\=\kill
{\bf Algorithm 1a: Open Chains}\\
{\sf repeat until} chain straightened {\sf do}\\
        \> {\sf 1: if} $s_g$ is {\em free\/} {\sf then}\\
        \> \> Construct obstruction diagram Ob$(v_0)$ on $3$-sphere.\\
        \> \> Apply motion planning to move $v_0$ to $v_g$.\\
         \> {\sf 2: else} $s_g$ is {\em intersected\/}\\
	\> \> Construct obstruction diagram Ob$(v_1)$ on $2$-sphere.\\
        \> \> Move $v_1$ so that the goal position is not intersected.\\
\end{tabbing}
\end{minipage}
}
\end{center}

\subsubsection{Step~1: $s_g$ is free}
Our argument depends on some basic intersection facts, which we
formulate in $\R^d$ in a series of lemmas before specializing to
the $d=3$ and $d=4$ cases we need.

\paragraph{Geometric Intersections in $\R^d$.}
\seclab{Geometric.Intersections}

Let the coordinates of $\R^d$ be 
$x_1,x_2,\ldots,x_d$.
A {\em $k$-flat\/} is 
the translate of a subspace spanned by $k$ linearly independent vectors.
Flats for $k=0,1,2$ are also called points, lines, and planes.
A $k$-sphere 
is the set of points in a $(k+1)$-flat at a fixed radius from a
point (its {\em center\/}) in that flat.
A $0$-sphere is a set of two points,
a circle is a $1$-sphere, and the surface of a ball in $\R^3$ 
is a $2$-sphere.
When emphasizing the topology of a $k$-sphere, we will use
the symbol $\Sph^k$.

\begin{lemma}
The intersection of a $2$-flat $H$ (i.e., a plane)
with a $(d{-}1)$-sphere $S$ in
$\R^d$ is a circle, a point, or empty.
\lemlab{plane.sphere}
\end{lemma}
\begin{pf}
Translate and rotate the sphere and plane so that the sphere is centered
on the origin, and the plane is parallel to the $x_1x_2$-plane.
The equations of the sphere $S$ and the plane $H$ are then:
\begin{eqnarray}
S & : & x_1^2 + x_2^2 + \cdots + x_d^2 = r^2 \\
H & : & x_3 = a_3 \,,\, x_4 = a_4 \,,\, \cdots ,\, x_d=a_d
\end{eqnarray}
where the $a_i$ are constants.  Let $A^2 = \sum_{i=3}^d a_i^2$.
Then
\begin{eqnarray}
S \cap H & : & x_1^2 + x_2^2 + A^2 = r^2 \\
           &   & x_1^2 + x_2^2 = r^2 - A^2
\end{eqnarray}
If $r^2 < A^2$, the intersection is empty.
If $r^2 = A^2$, the intersection is the point $(0,0,a_3,\ldots,a_d)$.
If $r^2 > A^2$, the intersection is a circle in $H$
with radius $\sqrt{r^2 - A^2}$, and center $(0,0,a_3,\ldots,a_d)$.
\end{pf}

\begin{lemma}
The intersection of a (1D) line, ray, or segment
with a $(d{-}1)$-sphere $S$ in
$\R^d$ is at most two points,
i.e., it either contains one or two points or
is the empty set.
\lemlab{line.sphere}
\end{lemma}
\begin{pf}
Let $s = ab$ be a segment, and let the sphere center be $c$.
Let $H$ be the 2D plane determined by the three points
$a,b,c$, i.e., $H$ is the affine span of $\{a,b,c\}$.
Because $s \subset H$, we must have $s = s \cap H$.
So
\begin{eqnarray}
s \cap S & = & (s \cap H) \cap S \\
         & = & s \cap (H \cap S)
\end{eqnarray}
By Lemma~\lemref{plane.sphere}, $H \cap S$ is a circle,
and the claim for segments follows because a segment intersects
a circle in at most two points.
Rays and lines yield the same result by selecting $a$ and $b$ sufficiently
large.
\end{pf}


Let $a$, $b$, and $c$ be three distinct points in $\R^d$,
such that $c$ does not lie on the segment $ab$.
Call the set of points that lie on rays that start at $c$
and pass through a point of $ab$
a {\em triangle cone\/} $\tri_c(a,b)$.
If $(a,b,c)$ are collinear,
the triangle cone degenerates to a ray.

\begin{lemma}
The intersection of a triangle cone $\tri_c(a,b)$
with a $(d{-}1)$-sphere $S$ in
$\R^d$
consists of at most two connected components---and,
if $c$ is the center of $S$, 
of at most one component---each of which is a circular arc or a point.
\lemlab{tricone.sphere}
\end{lemma}
\begin{pf}
Let $\tri = \tri_c(a,b)$, 
and let $H$ be the 2D plane containing $\tri$.
Because $\tri \subset H$, $\tri = \tri \cap H$.
So $\tri \cap S = \tri \cap (H \cap S)$.
By Lemma~\lemref{plane.sphere}, $H \cap S$ is a circle $C$ in the
plane containing $\tri$.
So the problem reduces to the intersection of a triangle cone with
a circle.
As illustrated in Fig.~\figref{tricone}a,
this intersection is at most one arc if the cone's apex $c$ is
at the center of the $C$ ($\tri_1$ in the figure),
and at most two arcs otherwise ($\tri_2$ in the figure).
Any of the arcs illustrated could degenerate to points if
the cone is a ray.
(When $c$ is not the center of $S$, the arc could be the whole
circle $C$.)
\end{pf}
\begin{figure}[htbp]
\centering
\includegraphics[width=0.9\linewidth]{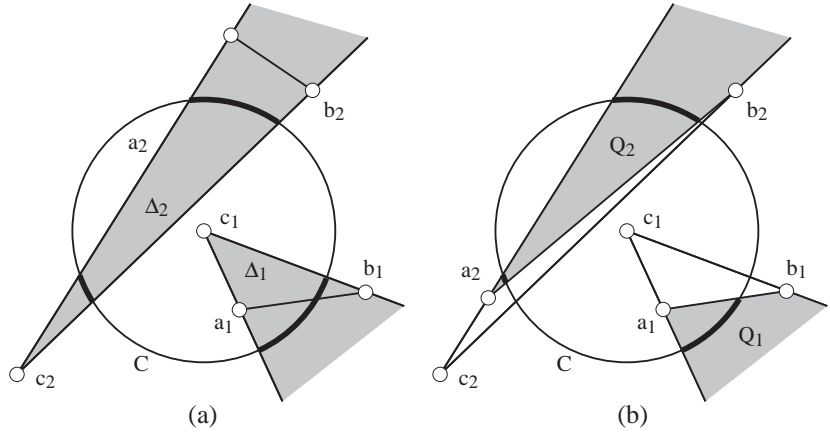}
\caption{(a) Intersections of triangle cones 
$\tri_1=\tri_{c_1}(a_1,b_1)$
and
$\tri_2=\tri_{c_2}(a_2,b_2)$
with a circle $C$ centered at $c_1$;
(b) Intersections of quadrilateral cones $Q_1$ and $Q_2$ with $C$.}
\figlab{tricone}
\end{figure}

We will need a slight extension of this lemma.
Define a {\em quadrilateral cone\/} $Q_c(a,b)$ to be
the closure of $\tri_c(a,b) \setminus t$,
where $t$ is the triangle determined by $(a,b,c)$.
Thus $Q_c(a,b)$ is all the points on the rays from $c$ at or beyond
$ab$. The next lemma says that the conclusion of the previous
lemma holds for quadrilateral cones as well.
\begin{lemma}
The intersection of a quadrilateral cone $Q_c(a,b)$
with a $(d{-}1)$-sphere $S$ in
$\R^d$
consists of at most two connected components---and,
if $c$ is the center of $S$, 
of at most one component---each of which is a circular arc or a point.
\lemlab{quadcone.sphere}
\end{lemma}
\begin{pf}
As Fig.~\figref{tricone}b makes clear, $Q_c(a,b)$
is just $\tri_c(a,b)$ intersected with a closed
halfplane in $H$ containing $ab$.
Intersecting the components from Lemma~\lemref{tricone.sphere}
with a halfplane cannot increase their number, and so the claim follows.
\end{pf}

\paragraph{Obstruction Diagram Ob$(v_0)$.}

Let $\C_0$ be the {\em configuration space\/} for vertex $v_0$
when $v_1$ is fixed: the set of all possible positions for $v_0$
that preserve the length of $v_1 v_0$.
$\C_0$ is a $3$-sphere $S$ in $\R^4$ centered at
$v_1$ with radius $\ell_0$.
Let $\F_0$ be the {\em free space\/} for vertex $v_0$
with all other vertices $v_i$ of the chain fixed:
the subset of $\C_0$ for which the chain is simple, i.e.,
for which $s_0$ does not intersect $s_i$, $i > 1$,
and $s_0$ intersects $s_1$ only at $v_1$.
We define the {\em obstruction diagram\/} $\B0$ for $v_0$
as the set such that $\F_0 = \C_0 \setminus \B0$.
Our goal is to describe, and ultimately construct, $\B0$.

To ease notation, let
$_j\tri_i=\tri_{v_j}(v_i,v_{i+1})$ be the triangle cone
with apex $v_j$ determined by segment $i$, and define
${_jQ_i} \subseteq {_j\tri_i}$ as the similar quadrilateral cone.



\begin{lemma}
The set of points $\B0 \subset \C_0$ 
in the $3$-sphere $S$
consists of at most $n-1$ components,
each of which is a circular arc of a circle or a point.
\lemlab{Obv0}
\end{lemma}
\begin{pf}
$\B0$ is the union of the obstructions contributed by
each segment $s_i$, $i > 1$,
plus the single point disallowing overlap with $s_1$.
If $s_0$ intersects $s_i$, then $v_0$ lies in the set
${_1Q_i}$ in $\R^4$,
for then $v_0$ lies on a ray from $v_1$ along $s_0$, beyond the
crossing with $s_i$.
(For example, in Fig.~\figref{tricone}b, we have
$c_1 = v_1$, $a_1 = v_i$, and $b_1=v_{i+1}$.)
Thus ${_1Q_i} \cap S$ is precisely the locus of positions of $v_0$
for which $s_0$ intersects $s_i$.
By Lemma~\lemref{quadcone.sphere}, this intersection is a circular arc
or a point.
Unioning over all $i>1$ establishes the claim.
\end{pf}

\noindent
This lemma is now immediate:

\begin{lemma}
If $v_0$'s goal position $v_g$ is free, then
$v_1$ may be straightened.
\lemlab{free.straighten}
\end{lemma}
\begin{pf}
Because $v_g$ is free, $v_g \not\in {\rm Ob}(v_0)$.
Because the given chain is assumed simple, the initial position
$v_0 \not\in {\rm Ob}(v_0)$.
The locus of possible $v_0$ positions
forms 
the $3$-sphere $S$.
The obstacles $\B0$ are a finite set of circular arcs and points.
The removal of $\B0$ from $S^3$ cannot disconnect $v_0$ from $v_g$.
This follows from the fact that $\R^d$ cannot be separated by
a subset of dimension of less than or equal to 
$d{-}2$~\cite[Thm.~3-61, p.~148]{hy-t-61}.
Neither then can $\Sph^d$ be so disconnected.
For suppose set $X$ disconnects two points $p$ and $q$ of $\Sph^d$.
Then stereographically project $\Sph^d$ to $\R^d$, from a center
not in $X$ or at the two points.  This produces a set $X'$ that
disconnects $p'$ from $q'$ in $\R^d$, contradicting the quoted theorem.

Therefore there is a path in $\F_0 = S \setminus \B0$ 
from $v_0$ to $v_g$,
which represents a continuous motion of $s_0$ that straightens $v_1$.
\end{pf}

It is this lemma which justifies the claim made in 
Section~\secref{Background}
that there can be no cages in 4D.
We will defer to Section~\secref{Canny.1} construction of the path
guaranteed by this lemma.

\subsubsection{Step~2: $s_g$ is intersected}
If $s_g$ is intersected, then rotating $s_0$ to the goal position
necessarily violates simplicity at the goal position.  In this case,
we slightly move $v_1$, the joint between $s_0$ and $s_1$, so that the
new goal position $s'_g$ is no longer intersected.


\noindent
That we can ``break'' the degeneracy of an intersected goal is
established by this lemma:
\begin{lemma}
$v_1$ may be moved to $v'_1$ while keeping all other vertices
fixed, so that the chain remains simple, and the new goal
$s'_g$ is not intersected.
\lemlab{inter.break}
\end{lemma}
\begin{pf}
Fix the positions of $v_0, v_2, v_3, \ldots, v_n$.
The $2$-sphere 
$$S = \{z \in \R^4 : |z-v_0|=\ell_0, |z-v_2|=\ell_1\}$$
represents all the possible positions for $v_1$ that
preserve the lengths of its incident links.
Note that $S$ consists of the intersection of two $3$-spheres.
Because we may assume that
the angle at 
$v_1$ is not already straightened, 
$S$ does not degenerate to a single point.
Thus $S$ is a $2$-sphere.


Now we construct an obstruction diagram 
$\D1$ on $S$ that is a superset of all those positions 
of $v_1$ for which~(1) the goal position $s_g$ (of $s_0$) is intersected,
or for which~(2) the chain $(v_0,v_1,v_2)$ intersects the remaining,
fixed chain $(v_2,\ldots,v_n)$.
We construct a superset rather than the precise obstruction set
because the former is easier but equally effective computationally.
\begin{enumerate}
\item Intersected goal positions $s_g$.
A goal segment $s_g$ lies on the ray from $v_2$ through $v_1$,
for it is exactly those $s_g$ that are straight at $v_1$.
For $s_g$ to intersect $s_i$, $v_1$ must lie in
$_2\tri_i$, the triangle cone with apex at $v_2$
and delimited by $s_i$.
See Fig.~\figref{sgint}.
\begin{figure}[htbp]
\centering
\includegraphics[height=7cm]{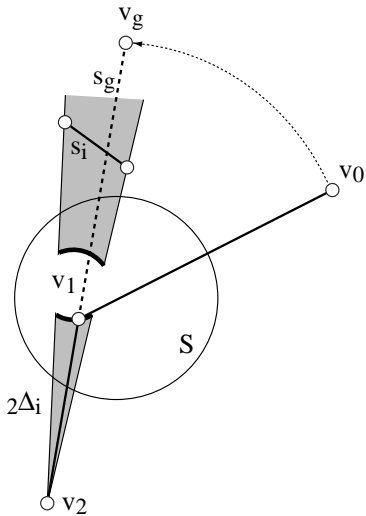}
\caption{The triangle cone $_2\tri_i$ intersects the sphere $S$
in at most two circular arcs.}
\figlab{sgint}
\end{figure}
Not every $v_1 \in {_2\tri_i}$ leads to intersection of $s_g$ with $s_i$:
$s_g$ must reach $s_i$.  The relevant subset of ${_2\tri_i}$ could
be detailed, but because it has one curved edge, we content ourselves
with a supset of the obstructions by forbidding $v_1$ anywhere
in ${_2\tri_i}$.

Applying Lemma~\lemref{tricone.sphere} shows that
$S \cap {_2\tri_i}$ contributes at most two arcs or points
to $\D1$, for each $i \not\in \{0,1\}$.

\item Intersections between $s_0$ and $s_1$ and the remainder of the chain.
$\D1$ also contains all the positions of $v_1$ that cause the
two adjacent links to intersect any of the other segments.
The link $v_2 v_1$ is clearly covered by $_2\tri_i$.
The link $v_0 v_1$ can be handled by the analogous triangle cone
$_0\tri_i$ with apex at $v_0$
and through $s_i$.
Again these sets provide a superset of the obstructions,
and 
Lemma~\lemref{tricone.sphere} again applies.
\end{enumerate}

Summing over all $i$ yields the obstruction superset $\D1$ composed of
at most $2 \cdot 3 (n-2) = O(n)$ arcs or points on $S$.
Thus $\D1$ is an arrangement of $O(n)$ arcs on a $2$-sphere,
with the initial position of $v_1$ lying on at least one arc
(because by hypothesis, $s_g$ is intersected).
Choosing any point $v'_1 \in S \setminus \D1$ interior to an
arrangement cell on whose boundary $v_1$ lies suffices to
establish the claim.
\end{pf}

Note that it is quite possible for $v_1$ to be confined
within a cell of the arrangement $\D1$, but that this
``cage'' is no impediment. We do not need a path 
from $v_1$ to an arbitrary point of $S$; rather we
only need a path to any unobstructed point $v'_1$.
Although we could construct the arrangement $\D1$
in $O(n^2\alpha(n))$ time and
$O(n^2)$ space~\cite{egppss-acptc-92,h-a-97},
for our limited goal of constructing just one point, we can do better:
\begin{lemma}
A move of $v_1$ to the position guaranteed by
Lemma~\lemref{inter.break} may be computed in
$O(n)$ time and $O(n)$ space.
\lemlab{inter.break.move}
\end{lemma}
\begin{pf}
Let $Z = \{a_1,\ldots,a_m\}$ be the collection of arcs of $V$
that contain $v_1$.
$Z$ may be found by a brute force check of
each of the $O(n)$ arcs.
Pick two arcs $a_1$ and $a_j$ angularly consecutive about $v_1$.
This can be accomplished in $O(n)$ time by fixing $a_1$, and
letting $a_j$ be the arc that makes the smallest angle with $a_1$.
Let $a$ be a circular arc ray (i.e., a directed great circle
starting and ending at $v_1$) that bisects this angle; or if
$Z$ only contains one arc, let $a$ be orthogonal to it;
or if $Z$ only contains one point, let $a$ be any ray from $v_1$.

Intersect $a$ with every arc and point of $\D1$, again in $O(n)$ time.
Let $\d$ be the distance from $v_1$ along $a$ to
the closest intersection.
Finally, choose $v'_1$ as the point $\d/2$ along $a$.
This point is guaranteed to be off $\D1$, and therefore unobstructed.

Moving (in one move) $v_1$ to $v'_1$ establishes a new goal $s'_g$ 
that is not intersected.
\end{pf}

\subsubsection{Motion Planning}
\seclab{Canny.1}
Now that we know we can perform Step~2 of Algorithm~1a in
$O(n)$ time per iteration,
we return to finding a path through $S^3$ for $v_0$,
as guaranteed by Lemma~\lemref{free.straighten}.
Motion planning between two points of the $3$-sphere $\F$
may be achieved by any general motion planning
algorithm~\cite[Sec.~40.1.1]{s-amp-97}.
For example,
Canny's Roadmap algorithm achieves a
time and space complexity of $O(n^k \log n)$, where $n$ is the number
of obstacles, and $k$ the number of degrees of freedom
in the robot's placements.  In our case, $k=3$.
His algorithm produces a piecewise algebraic
path through $\F$, of $O(n^k)$ pieces.
Each piece constitutes a constant number of moves,
with the constant depending on the algebraic degree of the curves,
which is bounded as a function of $k$.
Therefore each joint straightening can be accomplished
in $O(n^3)$ moves.
Repeating the planning and straightening $n$ times
leads to $O(n^4)$ moves in $O(n^4 \log n)$ time.
In the next section we reduce the $O(n^3)$ moves per joint straightening
to just $3$ moves per straightening.

\subsection{Algorithm 1b}
We have now established Theorem~\theoref{open.4D}, but with weaker
complexity bounds than claimed.
It is not surprising that applying a general motion planning
algorithm is wasteful in our relatively simple situation.
In fact
a significant improvement over Algorithm~1a can be achieved by
switching attention from the absolute position of $v_0$, to the
direction in which $s_0$ rotates.
Let the vector along $s_0$ be $w_0 = v_0 - v_1$,
and similarly let $w_g = v_g - v_1$.
Let $w$ be the {\em goal direction\/}:
a unit vector orthogonal to $w_g$ that represents
the direction in which $w_0$ should be rotated to move
it to its goal position. 
See Fig.~\figref{goal.direction}.
\begin{figure}[htbp]
\centering
\includegraphics[height=2.5in]{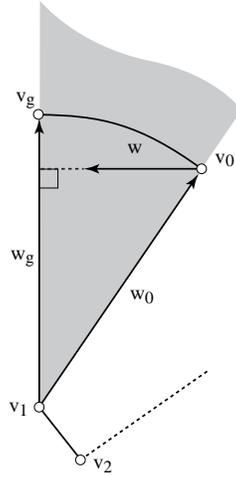}
\caption{The goal direction vector $w$ defines the direction
that $w_0$ should be rotated to reach $w_g$. The shaded triangle cone
${_1\tri(v_0,v_g)}$
is not crossed by any links of the chain if $w$ is unobstructed.}
\figlab{goal.direction}
\end{figure}
Thus $w$ is the unique unit vector 
pointing in the direction of the component of
$w_g - w_0$ orthogonal to $w_g$:
\begin{equation}
a_1 w_g + b_1 w  = w_g - w_0
\eqlab{goal.dir}
\end{equation}
for some reals $a_1 > 0$ and $b_1 > 0$. 
The space of possible directions $w$ forms a $2$-sphere rather than
the $3$-sphere we faced in Step~1 of Algorithm~1a.
This permits replacing the $O(n^3 \log n)$ moves per step from motion planning,
with at most two moves.
We now proceed to describe this.  
Because this represents a computational improvement only,
the proofs are only sketched.
More detailed proofs are contained in~\cite{Roxana}.

Algorithm~1b distinguishes three possibilities:
\begin{enumerate}
\item The goal position is {\em intersected\/} by some other
link of the chain (just as in Algorithm~1a).

\item The goal direction is {\em obstructed\/} in that
rotation of $s_0$ in the direction $w$ might hit some
link of the chain along its direct rotation to the goal position.
We again define a direction to be obstructed conservatively,
working with a superset of the true obstructions:
$w$ is obstructed if the triangular cone 
$\tri_{v_1}(v_0,v_g)={_1\tri(v_0,v_g)}$
is intersected by any $s_i$, $i > 1$.
\item The goal direction is {\em free\/}: it is not obstructed
(and so the goal position is not intersected).
\end{enumerate}
A high-level view of our second algorithm is as follows:
\begin{center}
\fbox{%
\begin{minipage}{56mm}
\begin{tabbing}
\hspace*{3mm}\=\hspace*{3mm}\=\hspace*{3mm}\=\kill
{\bf Algorithm 1b: Open Chains}\\
{\sf repeat until} chain straightened {\sf do}\\
	\> {\sf 1: if} $w$ is {\em free\/} {\sf then}\\
	\> \> Rotate $s_0$ directly to $s_g$.\\
	\> {\sf 2: else if} $w$ is {\em obstructed\/} {\sf then}\\
	\> \> Rotate $s_0$ to new position whose goal direction is free.\\
	\> {\sf 3: else if} $s_g$ is {\em intersected\/} {\sf then}\\
	\> \> Move $v_1$ so that the goal position is not intersected.
\end{tabbing}
\end{minipage}
}
\end{center}
Step~3 is identical to Step~2 of Algorithm~1a, so we only
discuss the first two steps.

\subsubsection{Step~1: $w$ is free}
By our definitions, $s_0$ may be rotated directly to $s_g$ without
hitting any other segment of the chain.  Because the goal
position $s_g$ is not intersected,
the chain remains
simple even after the rotation has been completed. Therefore, the link $s_0$
can be straightened in one move.

Note that this is the generic situation, in that for a ``random'' chain,
e.g., one whose vertex coordinates are chosen randomly from a 4D box,
each link can be straightened with Step~1 of the algorithm with
probability~$1$.  Steps~2 and~3 handle ``degenerate'' cases.
We exploit this in our implementation (Section~\secref{Implementation}).

\subsubsection{Step~2: $w$ is obstructed (but $s_g$ is not intersected)}

\paragraph{Detecting obstructions.}

When $w$ is obstructed, we again rely on construction of an
obstruction diagram.
First we describe the space in which the obstruction diagram
is embedded.

Consider the space of possible directions from which $s_0$
might approach $s_g$.
In 3D, this set of unit vectors forms a $1$-sphere, a circle, 
which can be viewed as orthogonal to $s_g$ and centered at $v_g$;
see Fig.~\figref{sphere.directions}a.
Similarly, in 4D, the set of possible approach directions toward $s_g$
forms a unit $2$-sphere $S$, which again we center on $v_g$.
Every point on this sphere represents a direction of approach
to $s_g$;
see Fig.~\figref{sphere.directions}b.

\begin{figure}[htbp]
\centering
\includegraphics[width=0.7\linewidth]{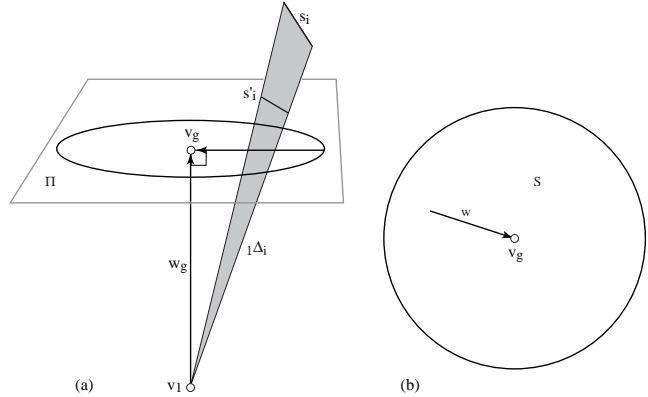}
\caption{(a) Directions approaching the goal position in 3D;
(b) $S$ is a $2$-sphere in $\R^4$.}
\figlab{sphere.directions}
\end{figure}

The {\em obstruction diagram\/} Ob$(s_g)$ is the set of vectors
$w$ representing obstructed goal directions for $s_g$.

\begin{lemma}
If the goal $s_g$ is not intersected,
the obstruction diagram Ob$(s_g)$ consists of at most $n$
arcs on $S$.
\lemlab{Ob}
\end{lemma}
\begin{pf}
Take an arbitrary segment $s_i$ of the chain, and ``project'' it 
to $s'_i$ in
the $3$-flat $\Pi \supset S$ orthogonal to $s_g$;
i.e., 
$s'_i = {_1\tri_i} \cap \Pi$.
See Fig.~\figref{sphere.directions}a for the 3D analog.
We first claim that the set of directions $w$ obstructed by
$s'_i$ is identical to those obstructed by $s_i$.
Next we determine this set of directions.
Every vector $w$ determined by a point on $S$ and its
center $v_g$, is
orthogonal to $s_g$ by our choice of $\Pi$.
So the set of $w$ obstructed by $s'_i$ is just those $w$
determined by the intersection of ${_g\tri}(s'_i)$ with $S$.
By Lemma~\lemref{tricone.sphere}, this is at most one arc on
the sphere.
See Fig.~\figref{sphere.4D}.
\begin{figure}[htbp]
\centering
\includegraphics[width=0.8\linewidth]{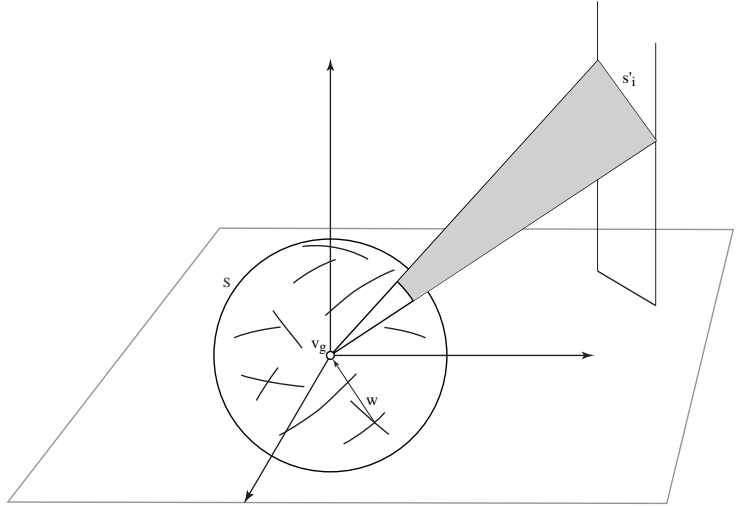}
\caption{In 4D, $s_i$ projects to $s_i'$ in the $3$-flat containing $S$,
and produces an arc of the obstruction diagram determined by the intersection
of the triangle cone ${_g\tri}(s'_i)$ with $S$.}
\figlab{sphere.4D}
\end{figure}
\end{pf}



Detection of obstruction therefore reduces to deciding if
$w$ lies on one or more arcs of an arrangement of circular arcs on
a $2$-sphere $S$, which can be accomplished in
$O(n)$ time and space as in
Lemma~\lemref{inter.break.move}.

\paragraph{Skirting obstructions.}

Our next task is to move $s_0$ when $w$ is obstructed so that
its new goal direction is free.
This task is similar to that handled in Lemma~\lemref{inter.break.move}---stepping off the arcs meeting at $w$---with 
one additional constraint: the move must maintain
the simplicity of the chain.
Note that Ob$(s_g)$ does not record chain simplicity, but rather
records free goal directions.  So we need to find
a $\Delta w$ that will
move $w$ to be free, while simultaneously maintaining
simplicity during the motion of $s_0$.
\begin{lemma}
If $w$ is obstructed, $s_0$ can be moved,
maintaining simplicity throughout,
so that its new goal direction $w' = w + \Delta w$ is unobstructed.
$\Delta w$ may be computed in $O(n)$ time and space.
\lemlab{Obs.avoidance}
\end{lemma}
\begin{pf}
%
Because the chain is initially simple, there must exist a $\b > 0$
such that rotation of $s_0$ about $v_1$ by an angle less than $\b$
leaves the chain simple.  This $\b$ can be computed by finding the
smallest distance $d$ from $s_0$ to any other segment, and using
the angle of a cone centered at $s_0$ of radius $d/2$.
Now $\Delta w$ is selected just as in Lemma~\lemref{inter.break.move},
but subject to this angle constraint.
\end{pf}

Note that because we have based our analysis on a fixed $s_g$,
moving $s_0$ does
not alter the obstruction diagram, which records obstructed directions of
approach to $s_g$.

\subsubsection{Algorithm~1b Complexity}
The algorithm straightens one joint in at most three moves:
one to move $v_1$ so the goal is not intersected (Step~3), one
to move $v_0$ so that the goal is not obstructed (Step~2),
and one to rotate directly to the goal (Step~1).
The total number of moves used by the algorithm is then
at most $3n = O(n)$.
For each of the $n$ iterations, Lemma~\lemref{Obs.avoidance}
shows that the computations can be performed in linear
time and space.
This then establishes the total time complexity of $O(n^2)$
claimed in Theorem~\theoref{open.4D}.
Because each move is performed independently, the obstruction
diagram arcs may be discarded after each iteration.
Thus the space requirements remain at $O(n)$.

\subsection{Implementation}
\seclab{Implementation}
We have implemented Algorithm~1b for chains in ``general position''
in C++.  
The program accepts a chain as input, and first checks if it is simple.  
If it is, the straightening
process starts; otherwise the program exits. 
The program then straightens the
chain link-by-link using Step~1, one move per link.
It also detects whether the goal is obstructed 
(Step~2) or intersected (Step~3) by solving sets of
linear equations,
but in those cases it simply halts; we have not implemented
the obstruction diagrams, or avoiding obstructions.
For a chain whose vertex coordinates are chosen randomly, the
program straightens it with probability~$1$, for then the degenerate
cases handled by Steps~2 and~3 (when a point,
$w$ or $v_1$, hits an arc on a $2$-sphere, 
e.g., Fig.~\figref{sphere.4D})   
are unlikely to occur.
The output of the program is a set of Geomview or Postscript files 
that animate the straightening process.
Fig.~\figref{chain.anim} shows 
output for a chain whose $n=100$ vertices were
chosen randomly and uniformly in $[0,1]^4$.
\begin{figure}[htbp]
\centering
\includegraphics[width=\linewidth]{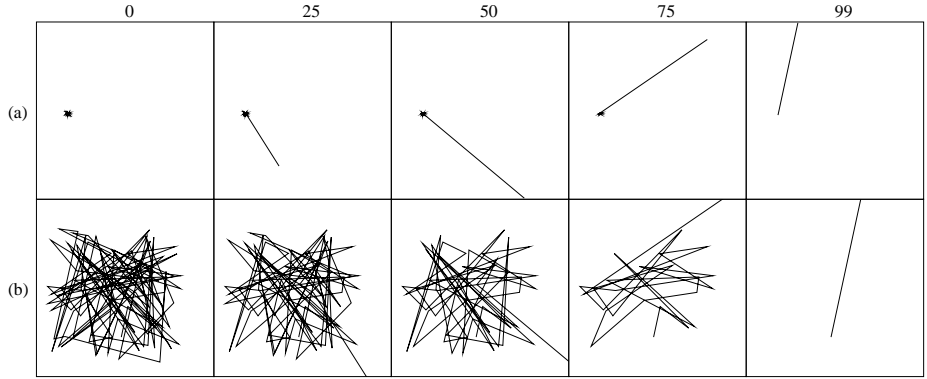}
\caption{Snapshots of the algorithm straightening a chain of $n=100$ vertices,
initially $(0)$, and after $25$, $50$, $75$, and all $99$ joints have been
straightened (left to right).
(a) Scale approx. 50:1; the entire chain is visible in each frame.
(b) Scale approx. 1:1; the straightened tail is ``off-screen.''
(The apparent link length changes are an artifact of the orthographic
projection of the 4D chain down to 2D.)}
\figlab{chain.anim}
\end{figure}

\section{Straightening Trees in 4D}
\seclab{Trees}
It will come as no surprise that essentially the same algorithm as
just described can straighten trees in 4D.
The reason is that each segment was considered a fixed obstruction
in the chain straightening algorithm, and whether those segments
form a chain or a tree is largely irrelevant, as long
as there is a free end. 
There is one spot
at which the difference between
a chain and a tree does matter, however: freeing up
an intersected goal position.
We concentrate on this difference in the description below.

\begin{figure}[htbp]
\centering
\includegraphics[width=0.7\linewidth]{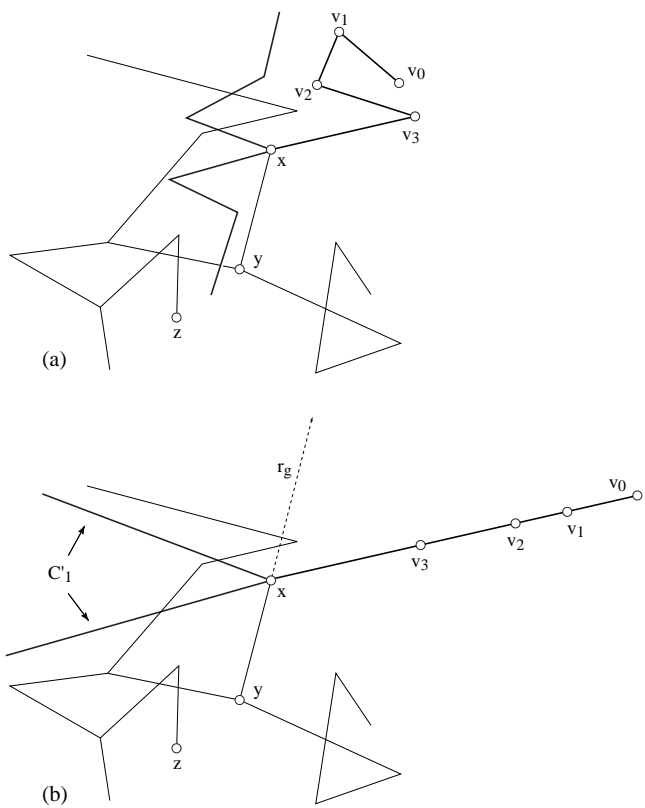}
\caption{(a) Tree $T$ rooted at $z$; (b) After straightening chains $C$
incident to $x$; $C'_1$ is the set of straightened chains excluding
one distinguished chain $(v_0, v_1, \ldots)$.}
\figlab{tree}
\end{figure}

\begin{center}
\fbox{%
\begin{minipage}{56mm}
\begin{tabbing}
\hspace*{3mm}\=\hspace*{3mm}\=\hspace*{3mm}\=\kill
{\bf Algorithm 2: Trees}\\
{\sf repeat} until straightened {\sf do}\\
\> {\sf 1:} Identify a node $x$ with chain descendants $C$.\\
\> {\sf 2:} Straighten each chain in $C$, forming $C'$.\\
\> {\sf 3: if} $r_g$ is intersected {\sf then}\\
\> \> Construct obstruction diagram Ob$(x)$ on 2-sphere.\\
\> \> Move $x$ so that $r_g$ not intersected.\\
\> {\sf 4:} Rotate each segment in $C'$ to $r_g$ and coalesce.
\end{tabbing}
\end{minipage}
}
\end{center}

Algorithm~2 chooses a leaf $z$ of the given tree $T$
as root, and then identifies some node $x$ all of whose
descendant subtrees are chains (Step~1). Call this set $C$;
see Fig.~\figref{tree}a.
Each chain in $C$ can be straightened one at a time via Algorithm~1,
leaving a set of straightened chains, or segments, $C'$ (Step~2).
Define the goal ray to be the extension of the parent segment
$yx$ incident to $x$; see Fig.~\figref{tree}b.
If $r_g$ is not intersected by any segment of $T \setminus C'$,
then each segment in $C'$ can be rotated to $r_g$,
each lying on top of one another (Step~4).  We can view them as coalesced
into a single link, reducing the degree of $x$ to $2$.  The
process then repeats.

If, however, $r_g$ is intersected (Step~3), we need to move $x$ so that the
goal ray becomes free.  There are several ways to achieve this;
we choose to parallel Step~2 of Algorithm~1a.  Let
$(v_0,v_1, \ldots, v_m)$ be one of the chains of $C'$, with
$v_m$ adjacent to $x$.  We distinguish this chain
from the others in $C'$; call the set of others $C'_1$.
Let the $2$-link chain $(v_0, x, y)$ play the role of
$(v_0, v_1, v_2)$ in Algorithm~1a.  In that algorithm we
argued that Ob$(v_1)$ is a set of arcs and points on a 
$2$-sphere (Fig.~\figref{sgint}).
Here will will reach the same conclusion for Ob$(x)$ on
the $2$-sphere $S$ of positions for $x$.

The only difference is that in the current situation, the
{\em star\/} of segments $C'_1$ is attached to $x$, and we
need to augment  Ob$(x)$ to reflect its obstructions.
We opt to translate $C'_1$ as $x$ moves; this gives rise
to two sets of constraints: (1)~those caused by a segment in
$C'_1$ intersecting a segment of 
$T' = T \setminus \{ C'_1 \cup xy \cup xv_0 \}$;
(2)~those caused by $xy$ or $xv_0$ intersecting a segment in $C'_1$.
For the first, the locus of positions of $x$ that cause some
$s \in C'_1$ to intersect some $s_i \in T'$ is a parallelogram,
congruent to the Minkowski sum $s \oplus s_i$.
Analogous to Lemma~\lemref{tricone.sphere},
it is easy to see that this holds:
\begin{lemma}
The intersection of a parallelogram
with a $(d{-}1)$-sphere $S$ in
$\R^d$
consists of at most four connected components,
each of which is an arc or a point.
\lemlab{para.sphere}
\end{lemma}
\noproof

Thus the constraints~(1) add $O(n)$ arcs or points to Ob$(x)$.
Constraints~(2) can be seen to consist of $O(n)$ points
on $S$: translating
the star $C'_1$ to $y$ determines the rays that $xy$ might
align with to cause $xy$ to intersect $C'_1$; and similarly
translating $C'_1$ to $v_0$ determines rays for intersection
with $x v_0$.  The two placements of $C'_1$ therefore generate
$O(n)$ additional point obstructions.

With Ob$(x)$ again a set of $O(n)$ arcs and points on a $2$-sphere,
Lemmas~\lemref{inter.break} and~\lemref{inter.break.move} hold,
leading to the same time complexities clamed for Algorithm~1,
and establishing Theorem~\theoref{tree.4D}.

\section{Convexifying Closed Chains in 4D}
\seclab{Closed}
Our algorithm for convexifying closed chains employs the
{\em line tracking\/} motions introduced in~\cite{lw-rcpce-95}.
Indeed our algorithm mimics theirs in that we repeatedly
apply line tracking motions, each of which straightens at least one
joint, until a triangle is obtained (which is a planar
convex polygon, as desired).
Although the overall design of our algorithm is identical,
the details are quite different, for there is a major
difference with~\cite{lw-rcpce-95}:
They permitted
self-intersections of the chain, whereas we do not.
This greatly complicates our task.%
\footnote{
	An alternative convexifying algorithm, again permitting
	self-intersections, is described
	in~\cite{s-scsc-73}.  Sallee accomplishes the same
	result by a different basic motion, involving four
	consecutive vertices rather than the five used in~\cite{lw-rcpce-95}.
}

Let $(v_0,v_1,v_2,v_3,v_4)$ be five consecutive vertices of
a closed polygonal chain.  We allow $v_0 = v_4$.
A {\em line tracking\/} motion of $v_2$ moves $v_2$ along some
line $L$ in space, while keeping both $v_0$ and $v_4$ fixed.
As long as the angle at joints $v_1$ and $v_3$
(the {\em elbows\/}) are neither $\pi$ (straight)
nor $0$ (folded), such a motion is possible.  Neither angle can
be $0$ because that would violate the simplicity of the chain.
Straightening one joint is precisely our goal, so we assume that
neither joint is straight; and therefore a line tracking motion
is possible.

We will choose $L$ and a direction along it so that the movement
increases the distance from $v_2$ to both $v_0$ and $v_4$
simultaneously.  This necessarily opens both elbow angles.
The motion stops when one elbow straightens.
The only issue is whether this can be done while maintaining simplicity.
Our aim is to prove this theorem:

\begin{theorem}
For a simple 4D chain  $(v_0,\ldots,v_4)$,
there exists a line tracking motion of $v_2$
that straightens either $v_1$ or $v_3$ (or both)
while maintaining simplicity
of the chain throughout the motion.
\theolab{line.tracking}
\end{theorem}

A high-level view of the algorithm is as follows:
\begin{center}
\fbox{%
\begin{minipage}{56mm}
\begin{tabbing}
\hspace*{3mm}\=\hspace*{3mm}\=\hspace*{3mm}\=\kill
{\bf Algorithm 3: Closed Chains}\\
{\sf repeat until} chain is a triangle {\sf do}\\
	\> Compute a line $L$ along which to move $v_2$.\\
	\> Compute free paths $\pi_1$ and $\pi_3$ for $v_1$ and $v_3$.\\
	\> Move $v_2$ along $L$, $v_1$ along $\pi_1$, and $v_2$ along $\pi_2$.\\
	\> Freeze the straightened joint $v_1$ or $v_3$.
\end{tabbing}
\end{minipage}
}
\end{center}

\subsection{Choosing $L$}
To fix $L$, the ray along which $v_2$ moves, we choose a
point $q \in \R^4$
different from $v_2$, and let $L$ be the ray from $v_2$ that
contains $v_2q$.
We will choose $q$
so that it is itself the point where one of the two joints
$v_1$ or $v_3$ becomes
straight while moving $v_2$ along $L$.

\begin{lemma}
A point $q$ determining an appropriate $L$ may always be found,
and in time and space $O(n^4)$.
\lemlab{Choosing.L}
\end{lemma}
\begin{pf}
We choose $q$ so that it satisfies these conditions:
\begin{enumerate}
\item Moving $v_2$ along $L$ increases the distance from $v_2$ to $v_0$ 
and to $v_4$.
\item Either $v_1$ or $v_3$ becomes straight, i.e.,
 $|qv_0|=|v_0v_1|+|v_1v_2|=r_0$, or $|qv_4|=|v_2v_3|+|v_3v_4|=r_4$
\item 
\begin{enumerate}
\item If $|qv_0|=r_0$, then $qv_0$ does not intersect
any other segment of the chain than those to which it is incident.
\item 
If $|qv_4|=r_4$, then $qv_4$ does not intersect
any other segment of the chain than those to which it is incident.
\end{enumerate}
\item
$v_2 q$ does not intersect a segment $s_i$, $i > 4$.
\end{enumerate}
Condition~3 ensures that our ``goal'' is not itself intersected,
in the sense used in Section~\secref{Open}.

Let $R_i$ be the set of points (the ``region'')
of $\R^4$ that satisfy Condition~$i$ above.
$R_1$ is the intersection of two
closed half-spaces containing $v_2$, orthogonal to $v_0v_2$ and $v_2v_4$
respectively.  Note that $v_2 \in R_1$.
If $v_0v_2$ and $v_2v_4$ lie on the same line, $R_1$ degenerates
to a $3$-flat orthogonal to that line; otherwise it is a 
$4$-dimensional set.\footnote{
	Although we could remove this possible degeneracy by moving $v_2$ in a
	neighborhood (while preserving simplicity) to break the collinearity,
	this is not necessary, as the proof goes through regardless.
}
See Fig.~\figref{L} for a lower dimensional analog of the situation.
\begin{figure}[htbp]

\centering
\includegraphics[width=0.6\linewidth]{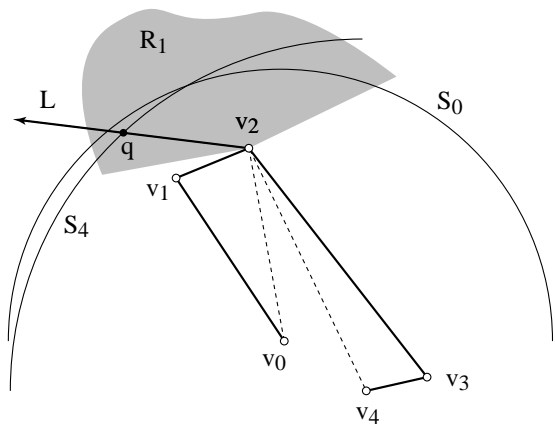}

\caption{Choosing $q \in L$. $R_1 \cap R_2 = R_1 \cap (S_0 \cup S_4)$.}
\figlab{L}
\end{figure}

The set of points $R_2 = S_0 \cup S_4$ 
in 4D that satisfy Condition~2 is the union of two
3-spheres, $S_0$ and $S_4$, centered at $v_0$ and $v_4$ 
and of radius
$r_0$ and $r_4$, respectively.
Because $|v_0v_2|<r_0$, $v_2$ is inside the $4$-ball bounded
by $S_0$.  Therefore, $R_1 \cap S_0 \neq \emptyset$.
Similarly, $R_1 \R^4 \setminus \cap S_4 \neq \emptyset$.
So 
$R_1 \cap R_2 \neq \emptyset$.
The dimensionality of this set depends on whether or
not $\{v_0, v_2, v_4\}$ are collinear:
if they are, the $3$-spheres are intersected by a $3$-flat
producing $2$-spheres;
if they are not, the $3$-spheres are intersected by a
$4$-dimensional wedge, producing $3$-dimensional regions of
the $3$-spheres.

Consider Condition~3a; clearly 3b is analogous.
We want all those points $q$ such that
$qv_0$ does not intersect any other link of the chain.
Clearly the points forbidden by segment $s_i$
lie in the triangle cone ${_0\tri_i}=\tri_{v_0}(v_i,v_{i+1})$,
just as in the proof of Lemma~\lemref{inter.break}.
Intersecting
${_0\tri_i}$ for all $i$ with $R_1 \cap R_2$ marks the set of
points that must be avoided in our choice of $q$:
$R_3 \supset \R^4 \setminus \bigcup_i {_0\tri_i}$.
It is easiest to concentrate on the intersection of
$_0\tri_i$ with the spheres in $R_2$.
By Lemma~\lemref{tricone.sphere}, we know this intersection
is at most two arcs or points, independent of the dimension of the
spheres.  So whether or not $\{v_0, v_2, v_4\}$ are collinear,
the intersection produces $O(n)$ arcs or points.
Similarly, Condition~4 leads to
$R_4 \supset \R^4 \setminus \bigcup_{i>4} {_2\tri_i}$,
for $v_2q$ can intersect $s_i$ only if $q$ lies in ${_2\tri_i}$.
Again, $O(n)$ arcs or points need be avoided in $R_1 \cap R_2$.
No union of arcs and points
can cover
the set $R_1 \cap R_2$, which is either $2$- or $3$-dimensional.
Thus $\bigcap_i R_i \neq \emptyset$.
We need only choose a $q$ in this set.

There are a variety of ways to choose such a $q$ algorithmically.
A naive method is to first
construct an arrangement of $2$-flats in $\R^4$
each containing a triangle $_0\tri_i$ or $_2\tri_i$.
This computation could be performed in $O(n^4)$ time and space
\cite{ess-ztha-93}.
Intersecting this arrangement with
the halfspaces delimiting $R_1$ and the $3$-spheres
$S_0$ and $S_4$ leave us cells bound by algebraic surfaces
inside $\bigcap_i R_i$.  The centroid of any such cell
can be selected as $q$. 
\end{pf}

\subsection{Line Tracking in 3D}
We start by thinking about the analogous situation in 3D.
This will both set notation, and ground intuition by showing
why Theorem~\theoref{line.tracking} does not hold in 3D.

\subsubsection{Topology of Configuration Space in 3D}
Let $\R_{[0,1)}$ be the interval $[0,1)$ on the real line, open at $1$.
We will parametrize the location of $v_2$ along $L$ by $t \in [0,1)$,
with $t=0$ the start, and $t=1$ when $v_2$ reaches the
$q$ of Lemma~\lemref{Choosing.L},
the first time
at which a joint, 
straightens. Let this joint
be $v_1$ without loss of
generality.
Let $\C'$ be the configuration space of the four-link system
in isolation, permitting intersections between the links,
the prime to remind us that $t=1$ has been excluded.
We claim that
\begin{equation}
\C' = \Sph^1 \times \Sph^1 \times \R_{[0,1)} .
\eqlab{C3}
\end{equation}
This can be seen as follows.
Fix some $t$ so that $v_2$ is fixed.  Then each of $v_1$ and $v_3$
is free to rotate (independently) on a circle in $\R^3$ 
centered on the axis $v_0 v_2$ and
$v_2 v_4$ respectively.
As $t$ varies from $0$ to $1$, these circles move in space,
and grow and shrink in radius;
see Fig.~\figref{circles.3D}.
\begin{figure}[htbp]

\centering
\includegraphics[height=4.5cm]{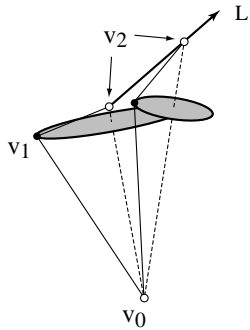}
\caption{In 3D, the circle on which $v_1$ may lie moves in space
as $v_2$ slides up $L$.}
\figlab{circles.3D}
\end{figure}
At $t=1$ the $v_1$ circle shrinks
to a point.  But for $t \in [0,1)$, both circles retain a positive
radius.  Thus the configuration space $\C$
has the topology of $\Sph^1 \times \Sph^1$ for each $t$,
and the claim follows.

\subsubsection{Obstruction Diagram in 3D}
As in Section~\secref{Open}, we incorporate the obstacles representing
the other links
via an ``obstruction diagram.''
We start by ignoring the four moving links as obstructions,
and only consider the remaining, fixed links of the polygonal chain as
obstacles.
We develop the obstruction diagram first for fixed $t$, so that
the relevant configuration space is $\Sph^1 \times \Sph^1$.
Because we are ignoring the moving links as obstructions,
movement on the two circles is independent, so it suffices
to determine the obstruction diagram Ob$(v_1)$ on one $1$-sphere/circle
$S_1$, that for $v_1$.
The following lemma will be key in 4D:
\begin{lemma}
In 3D, if
$(v_2 - v_0) \cdot (v_1 - v_0) \neq 0$
and $(v_2 - v_0) \cdot (v_1 - v_2) \neq 0$,
then a single segment contributes at most four points to
Ob$(v_1)$.
Otherwise, if either dot product is zero,
a segment could obstruct a finite-length arc
of the $S_1$ circle for $v_1$.
\lemlab{obs.pts.3D}
\end{lemma}
\begin{pf}
We only sketch a proof, leaving details for the 4D case considered
below.  Spinning $v_1$ along its circle of freedom while maintaining
$v_0$ and $v_2$ fixed traces out a ``spindle'' shape,
which can be viewed as the union of two cones.  A segment $s$
that does not lie along a line through either $v_0$ or $v_2$
can intersect each cone in at most two points, and so intersect
the spindle in at most four points.  See Fig.~\figref{four.obstacles}.
\begin{figure}[htbp]
\begin{minipage}[b]{0.45\linewidth}
\centering
\includegraphics[width=0.95\linewidth]{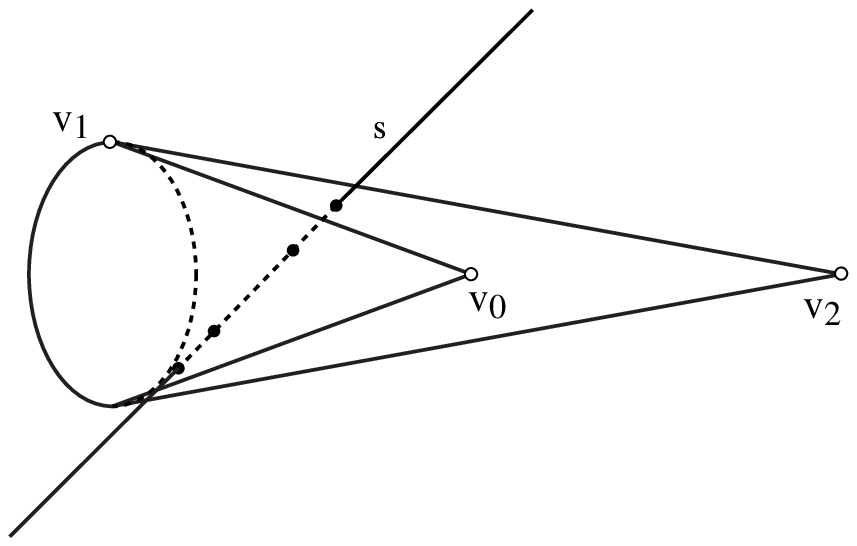}
\caption{One segment $s$ can contribute four points to Ob$(v_1)$.}
\figlab{four.obstacles}
\end{minipage}
\hspace{0.1\linewidth}
\begin{minipage}[b]{0.45\linewidth}
\centering
\includegraphics[width=0.5\linewidth]{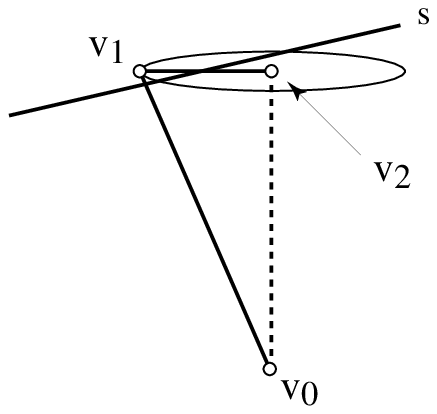}
\caption{$(v_2-v_0) \cdot (v_1-v_2)=0$ and segment $s$ (which lies in the
plane of the circle) contributes an arc to the obstruction diagram
Ob$(v_1)$.}
\figlab{dot.zero}
\end{minipage}
\end{figure}
These four segment-cone intersection points correspond one-to-one
with four $v_1$ positions on $S_1$ at which there is an intersection
between the $2$-link chain $(v_0,v_1,v_2)$ and $s$.

If the segment $s$ lies in the surface
of the cone, then it contributes just one point to the diagram,
corresponding to the angle of spin that aligns one of the two
links with the obstacle segment.

Finally, if either of the two links $v_0v_1$ or $v_1v_2$ is
orthogonal to the axis of the spindle, i.e., either
dot product is zero,
then a segment obstacle could obstruct the entire circle, for
one of the cones is then degenerately flat.
As Fig.~\figref{dot.zero} illustrates, here a segment might
obstruct a range of rotations of $v_1-v_2$, producing an arc in
Ob$(v_1)$.  
\end{pf}

\subsubsection{Disconnected Free Space in 3D}
Let $v_1(t)$ represent the position of $v_1$
on its circle $S_1$ at a particular time $t$.
The goal
is for the links $(v_0,v_1,v_2)$ to avoid all obstacles,
which means that $v_1(t)$ should avoid points of the obstruction
diagram. 
If we ignore for now the orthogonality case, then we have
the situation that a finite set of links produce an obstruction
diagram consisting of a finite set of points on $S_1$.
As $t$ moves, these points wander around the circle, disappear,
enter, join, or split.
The moving links, previously ignored,
just add a few more points to the obstruction
diagram, moving in a different manner.
The diagram for the configuration space for $v_1$ then looks like
arcs on the tube-like $\Sph \times \R_{[0,1)}$.
It is clear that it is possible for the point
$v_1(t)$ to be ``captured'' between two points of the
obstruction diagram which move together and squeeze
$v_1(t)$ into a collision.
See Fig.~\figref{config.tube}.  In this case, the free space
for the point $v_1$ is not connected from $p_1(0)$ to $p_1(1)$.
\begin{figure}[htbp]
\centering
\includegraphics[height=6cm]{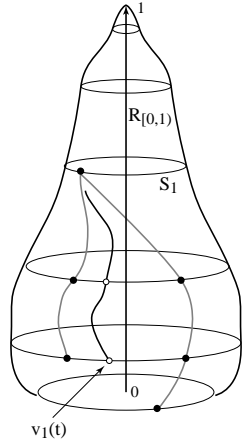}
\caption{Point $v_1(t)$ is ``captured'' by two obstacle
points in configuration space, the tube-like surface.}
\figlab{config.tube}
\end{figure}
And indeed it is easy to
``cage in'' the moving links by the fixed links so that no
straightening is possible.  Our next task is to show that
such caging-in is impossible in 4D.

\subsection{Line Tracking in 4D}
\subsubsection{Topology of Configuration Space in 4D}
Turning now to 4D, 
exactly analogous to the situation in 3D,
an elbow at the join of
two links has a space of possible motions in 4D that is topologically
$\Sph^2$, for it is the intersection of two 3-spheres.
Thus the configuration space $\C'$ of our four-link chain
for $t\in[0,1)$, ignoring
self-intersections, is
\begin{equation}
\C' = \Sph^2 \times \Sph^2 \times \R_{[0,1)} \;.
\eqlab{C5}
\end{equation}
At $t=1$ at least one of the $2$-spheres
shrinks to a point.

\subsubsection{Obstruction Diagram in 4D}
As in 3D, we analyze the obstruction diagram on
one $2$-sphere $S_1$, that for $v_1$, at a fixed value of $t$:
Ob$(v_1)$.
Let $v_1(t)$ represent the position of $v_1$
on its sphere $S_1$ at time $t$.
We seek the set of points Ob$(v_1)$ for which the
links $(v_0,v_1,v_2)$ intersect some other segment
of the chain, $s_4, s_5, \ldots, s_n$.
Just as in 3D, Ob$(v_1)$
is (in nondegenerate situations) a finite set of points.
This claim relies on how a line may intersect a cone.

Define a {\em $(d{-}1)$-cone\/} $C(a,b,\t)$, for 
apex point $a$, axis point $b$, and cone angle $\t \in [0,\pi/2]$,
to be the set of points $p \in \R^d$ that form an angle $\t$
with respect to the axis, i.e., which satisfy:
\begin{equation}
( p - a ) \cdot (b - a ) = |p - a| |b - a| \cos \t \;.
\eqlab{cone}
\end{equation}
For the extreme values of $\t$,
$C(a,b,0)$ is a ray from $a$ through $b$,
and $C(a,b,\pi/2)$ is a $(d{-}1)$-flat containing $a$ and orthogonal to $ab$.
Note that a $1$-cone is not the triangle cone from
Section~\secref{Geometric.Intersections}; rather a $1$-cone is the union
of two rays from $a$.
In 3D, $C(a,b,\t)$ is the surface of a right circular cone whose axis is the
ray from $a$ through $b$, and which form the angle $\t$ with the axis
at $a$ (cf.~Fig.~\figref{four.obstacles}).
Its intersection with a plane orthogonal to $ab$ is a circle.
In 4D,  $C(a,b,\t)$ is a ``right spherical cone,'' whose intersection
with a $3$-flat orthogonal to $ab$ is a $2$-sphere.
Note that it is no restriction to insist that $\t \in [0,\pi/2]$,
for we can ensure this for $\t > \pi/2$
by selecting
an axis point $b'$ for the cone to be on the other
side of the apex $a$, on the line containing $ab$,
thereby ``reflecting'' $\t$ to $\pi - \t$.

\begin{lemma}
The intersection of the $(d{-}1)$-cone $C(a,b,\t)$,
$\t \neq \pi/2$, with
a line, ray, or segment whose
containing line does not include the apex $a$,
is at most two points:  two points, one point, or empty.
\lemlab{line.cone}
\end{lemma}
This claim can be seen intuitively as follows.
Let $C$ be the cone and
$s$ a segment in $\R^d$.
If $s$ is contained in a $(d{-}1)$-flat $\Pi$ orthogonal to $ab$,
then because $\Pi \cap C$ is a sphere, the result follows
from Lemma~\lemref{line.sphere}.
Otherwise $s$ is contained in a flat whose intersection with $C$
is an ellipsoid, and the result follows because an ellipsoid is
affinely equivalent to a sphere~\cite[p.~95]{s-pg-88}.

\begin{pf}
Let $|ab| = 1$ without
loss of generality.  Translate and rotate $C$ so that
$a = (0,0,\ldots,0)$ and $b=(1,0,0,0,\ldots,0)$.
For a point $p=(x_1,\ldots,x_d)$, Eq.~\eqref{cone} reduces to
\begin{eqnarray}
p \cdot b & = & |p| \cos \t \\
(x_1,\ldots,x_d) \cdot (1,0,0,0,\ldots,0) & = & \sqrt{x_1^2+\cdots+x_d^2} \cos \t \\
x_1^2 & = & (x_1^2+\cdots+x_d^2) \cos^2 \t
\eqlab{cone.dot}
\end{eqnarray}
Represent the point $p$ via the parameter $t$:
\begin{equation}
p = (\a_1 + \b_1 t,\ldots,\a_d + \b_d t ) \; .
\end{equation}
Substitution of this into Eq.~\eqref{cone.dot} yields a quadratic
equation in $t$, which has at most two roots.

We now examine the degenerate solutions.
Because we assumed that $\t \neq \pi/2$, $\cos \t \neq 0$.
Thus the righthand side of Eq.~\eqref{cone.dot} can only be
zero when $x_1^2+\cdots+x_d^2 = 0$, i.e., when $p=(0,0,\ldots,0)$
is the apex $a$.  This corresponds to a line through $a$,
excluded by our assumptions.
\end{pf}

\begin{lemma}
In 4D, if
$(v_2 - v_0) \cdot (v_1 - v_0) \neq 0$
and $(v_2 - v_0) \cdot (v_1 - v_2) \neq 0$,
then a single segment $s$ contributes at most four points to
Ob$(v_1)$.
\lemlab{obs.4pts.4D}
\end{lemma}
\begin{pf}
Moving $v_1$ sweeps out two finite cones, which are truncations of
the infinite cones
$C(v_0, v_2, \t_0)$ and $C(v_2, v_0, \t_2)$,
with
\begin{eqnarray}
(v_2 - v_0) \cdot (v_1 - v_0) = |v_2 - v_0| |v_1 - v_0| \cos \t_0 
\eqlab{t0} \\
(v_2 - v_0) \cdot (v_1 - v_2) = |v_2 - v_0| |v_1 - v_2| \cos \t_2 
\end{eqnarray}
By the preconditions of the lemma, we have $\t_j \neq \pi/2$, $j=0,2$,
so we may assume $\t_j \in [0,\pi/2)$ by the reflection maneuver
suggested previously.
Consider two cases:
\begin{enumerate}
\item The line containing $s$ does not pass through either
cone apex, $v_0$ or $v_2$.
The conditions of Lemma~\lemref{line.cone} are satisfied, establishing
that $s$ intersects the two cones in at most four points.
Each of these points fixes a position of $v_1$ corresponding
to an obstruction, and so contributes this point to Ob$(v_1)$.
\item The line $H$ containing $s$ passes through $v_0$ 
(the case through $v_2$ is exactly analogous 
and will not be treated separately).
Then it may be that $s \cap C(v_0, v_2, \t_0)$ is a subsegment of $s$.
This is because the vector $p - v_0$ makes the same
angle with $v_2 - v_0$ for all $p \in s$ (cf.~Eq.~\eqref{cone}).
In this case, $s$ obstructs the unique position of $v_1$ that
places it on $H$, and so contributes just one point to Ob$(v_1)$.
Together with the at most two points from the other cone,
$s$ generates at most three points of Ob$(v_1)$.
\end{enumerate}
\end{pf}

The case excluded by the precondition of Lemma~\lemref{obs.4pts.4D} refers to
the situation in which one cone is degenerately flat,
as previously illustrated in Fig.~\figref{dot.zero}.  We now analyze
this situation in detail.
\begin{lemma}
If $(v_2 - v_0) \cdot (v_1 - v_0) = 0$,
then Ob$(v_1)$ is a finite set of points and arcs on $S_1$
(the $2$-sphere of $v_1$ positions).
\lemlab{cone.flat.4D}
\end{lemma}
\begin{pf}
In this case $\t_0 = \pi/2$ from Eq.~\eqref{t0},
and the infinite cone $C(v_0,v_2,\pi/2)$ degenerates to the 
$3$-flat orthogonal to the axis $v_0 v_2$ and including
the apex $v_0$.
The finite cone swept out by the link $s_0 = v_0 v_1$ is
a ball $B_0$ of radius $\ell_0$ centered at $v_0$.
In the 3D situation, $B_0$ is a disk
(cf.~Fig.~\figref{dot.zero}); in 4D, it is a solid sphere
whose boundary is a $2$-sphere $S_1$ representing the possible positions 
for $v_1$.

The obstructed positions on $S_1$  are those for which
$s_0$ intersects some segment $s_i$.  
Consider two possibilities:
\begin{enumerate}
\item $s_i$ does not lie in the same $3$-flat of $\R^4$ as $S_1$.
Then $s_i$ intersects $B_0$ in at most one point $p$ (because
it can intersect the flat in at most one point), and then only
when $s_0$ passes through $p$ do we have an obstruction.
Thus $s_i$ contributes one point to Ob$(v_1)$.
\item $s_i$ is in the same $3$-flat as $S_1$.
Now we have a situation exactly analogous to that shown in
Fig.~\figref{sphere.4D}:
the obstruction is the intersection of the triangle
cone ${_0\tri_i}$ with $S_1$.  
Lemma~\lemref{tricone.sphere} then establishes that
$s$ adds at most two arcs or points to Ob$(v_1)$.
\end{enumerate}
\end{pf}

\begin{lemma}
The condition $(v_2 - v_0) \cdot (v_1 - v_0) = 0$
can hold at most one value of
$t \in [0,1]$
during the movement of $v_2$ along $L$.
\lemlab{dotzero.once}
\end{lemma}
\begin{pf}
This follows immediately from our choice of $L$, which guarantees
that the distance $|v_0 v_2|$ increases, and so the angle at $v_1$ opens.
This angle can therefore pass through $\pi/2$ at most once.
See Fig.~\figref{dotzero.once}.
\begin{figure}[htbp]

\centering
\includegraphics[height=5.5cm]{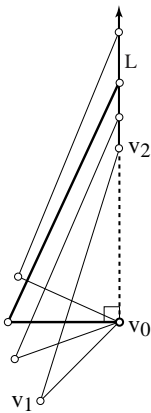}
\caption{The special condition $(v_2 - v_0) \cdot (v_1 - v_0) = 0$
holds at most once.}
\figlab{dotzero.once}
\end{figure}
\end{pf}

\subsubsection{Connected Free Space in 4D}
Again let $v_1(t)$ represent the position of $v_1$ on
its $2$-sphere $S_1$ of possible positions.
We first describe the free space for the motion of the
$2$-link chain $(v_0,v_1,v_2)$, avoiding the fixed
links $s_4, s_5, \ldots, s_n$.  
It is a subset of $\Sph^2 \times \R_{[0,1)}$.
For each $t \in [0,1)$, we know from Lemma~\lemref{obs.4pts.4D}
that Ob$(v_1)$ is a set of points or arcs; and
from Lemma~\lemref{dotzero.once} we know Ob$(v_1)$ is a finite set
of points, except for at most one $t$, at which it is a set of
points and arcs.
Thus if $v_1(t)$ avoids these obstructions, it avoids intersection
with the remainder of the chain.

But now it should be clear that it is easy for $v_1(t)$ to 
``run away'' from the obstructions.  Think of its sphere of possible
positions growing and shrinking with time $t$.  $v_1(t)$ must
avoid a set of points at any one time, and once
(cf.~Lemma~\lemref{dotzero.once}), a set of arcs.
This is easily done:  there is no way to ``cage'' in $v_1(t)$
with these obstacles.
Another view of this situation is that the configuration space
$\Sph^2 \times \R_{[0,1)}$ is $3$-dimensional, and the
obstructions Ob$(v_1(t))$ for $t \in [0,1)$ are $1$- or $0$-dimensional,
and the removal of a 1D set cannot disconnect a 3D set
(cf.~proof of Lemma~\lemref{free.straighten}).

The remainder of this subsection establishes this claim more formally.
A {\em path\/} in a topological space $X$ is a continuous function 
$\gamma :[0,1]\rightarrow X$.
A space is {\em path-connected\/} if any two of its 
points can be joined by a path~\cite{a-bt-79}.
We first work with the space $\C'_1$:
the positions for $v_1$, for $t \in [0,1)$.
Later we will add in $t=1$, and positions for $v_3$.
\begin{lemma}
The free space $\F'_1 \subset \C'_1$ for $v_1$ in the configuration space
$\C'_1 = \Sph^2 \times \R_{[0,1)}$ is path-connected.
\lemlab{connected.1}
\end{lemma}
\begin{pf}
It will help to view our configuration space as follows.
The $2$-sphere $S_1$ is represented by a flat
two-dimensional sheet, and
$\R_{[0,1)}$ is represented as a vertical axis.
The result is a three-dimensional space, analogous
to Fig.~\figref{config.tube}, that could look as
depicted in
Fig.~\figref{connected.space}.
The point obstacles Ob$(v_1)$ become paths monotone with respect
to the vertical $t$-axis.  At one $t=t_1$ we may have arc obstacles as well.
We need to show that $v_1(0)$ is connected by a path to $v_1(t')$,
for any $t' < 1$.
\begin{figure}[htbp]
\centering
\includegraphics[height=8cm]{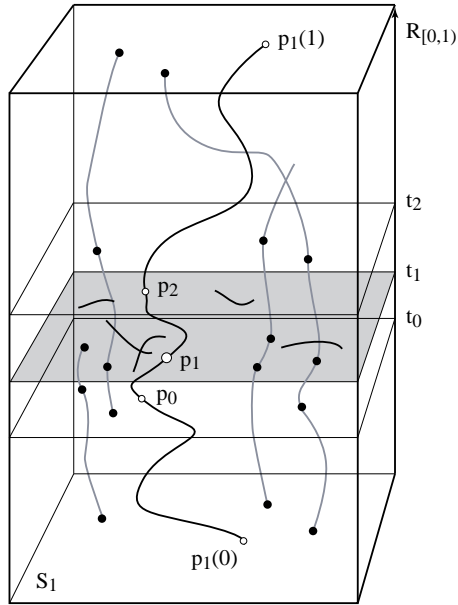}
\caption{The free space $\F_1$ for $v_1$ is path-connected.
$\pi_1$ (dark) connects $p_1(0)$ to $p_1(1)$.
Ob$(v_1)$ includes points at a fixed $t$,
forming curves (shaded) over time.
The shaded subspace at time $t=t_1$ includes arcs in Ob$(v_1)$.}
\figlab{connected.space}
\end{figure}
We proceed in two cases.
\begin{enumerate}
\item Ob$(v_1)$ contains only points for all $t \in [0,1)$.
Let $N$ be the maximum number of points in Ob$(v_1)$ over all $t$;
we know $N \le 2n$.  
A $2$-sphere with a finite number $N$ points removed is
path-connected.  For each $t$, remove $N$ points from
the corresponding $S_1(t)$: those in Ob$(v_1)$ at that $t$,
and extra distinct points to ``pad out'' to $N$.
Any two spheres with the same number of points removed are
homeomorphic.  Therefore $\F'_1$ is
homeomorphic to $S_1(0) \times \R_{[0,1)}$.
Because each of those spaces is path-connected, and the product
of two path-connected spaces is path-connected, we have
established the claim.
\item Ob$(v_1)$ contains arcs at $t=t_1$.
The main idea here is to choose a point $p_1 = v_1(t_1)$ that is unobstructed
at time $t=t_1$, and then connect from $v_1(0)$ to $p_1$, and from $p_1$
to $v_1(t')$.
It is clear, as we have shown in 
Case~1, that the spaces $\F_- = \C_{/t \in [0,t_1)}$
and $\F_+ = \F_{/t \in (t_1,1)}$ are path 
connected.
We will prove that there exist points $p_0 \in \F_-$, and 
$p_2 \in \F_+$ such that $p_0$ and $p_2$ are connected by a path.

We will call a point $p$ {\em free\/} if it does not belong to any obstruction
diagram. Let $p_1 \in S_1(t_1)$ be a free point on $S_1$ at $t$.
It is clear that such a point exists, since the obstruction diagram is a 
finite set of arcs and points. 
It is also clear that 
there exists a 
neighborhood $U \subset \F'_1$ of $p_1$ all of whose points are free.
Choose $p_0 \in U$, $p_0 \in S_1(t_0)$, $t_0 < t_1$ and  
$p_2 \in U$, $p_2 \in S_1(t_2)$, $t_1 < t_2$.
See Fig.~\figref{connected.space}.
Both points are free and can be connected by a path in $U$ to $p_1$.
But $p_0 \in \F_-$
and $p_2 \in \F_+$, both path connected spaces.
Thus we may connect $v_1(0)$ to $p_0$ to $p_1$ to $p_2$ to $v_1(t')$.
\end{enumerate}
\end{pf}

We now address the endpoint $t=1$, extending $C'_1$ to $C_1$
for $t \in [0,1]$.
As $v_2$ approaches $q$ on $L$, one of the spheres, that for $v_1$
by our assumptions,
shrinks to zero radius.
Thus Fig.~\figref{connected.space} is not an accurate depiction 
near $t=1$, for the configuration space narrows to a point here.
\begin{lemma}
The free space $\F_1$ for $v_1$ in the full configuration space
$\C_1$ is path-connected.
\lemlab{connected.2}
\end{lemma}
\begin{pf}
We have chosen $q$ and $L$ in Lemma~\lemref{Choosing.L}
so that the $t=1$ endpoint is free in the sense that the straightened
chain $v_0 v_1 v_2$ does not intersect the fixed portion of the chain.
Thus there is a neighborhood $U$ of $t=1$ such that $\C_1$ is
devoid of all obstructions within that neighborhood.
Choose $t' \in U$ and apply Lemma~\lemref{connected.1}
to yield a path from $v_1(0)$ to $v_1(t')$.
Connect within $U$ from $v_1(t')$ to the endpoint $v_1(1)$.
\end{pf}

Now we include $v_3$ in the analysis.
\begin{lemma}
The free space $\F \subset \C$ for both $v_1$ and $v_3$
in the configuration space
$\C$ for $t \in [0,1]$ is path-connected.
\lemlab{connected.3}
\end{lemma}
\begin{pf}
The key here is the independence of the motions of $v_1$ and $v_3$.
Let $\pi_1$ be a path for $v_1(t)$ through $\F_1$, whose
existence is guaranteed by 
Lemmas~\lemref{connected.1} and~\lemref{connected.2}.
Now construct $\F_3$ as the possible positions $v_3(t)$
for $v_3$, avoiding at each time Ob$(v_3(t))$, where this
time the obstructions include not only the fixed
links $s_4, s_5, \ldots, s_n$,
but also the two moving links $s_0$ and $s_1$, determined by $\pi_1$.
If $v_3(t)$ avoids Ob$(v_3(t))$ for each $t$, then all intersections
are avoided:
we do not need to include the moving links in $\F_1$, because
intersection is symmetric---if the links $s_2$ and $s_3$ do
not intersect $s_0$ and $s_1$, then $s_0$ and $s_1$ do not
intersect $s_2$ and $s_3$.

For a fixed $t$, the obstacles are fixed segments, and
Ob$(v_3)$ is again a finite set of points,
or, for at most one $t$, a set of arcs:
Lemmas~\lemref{obs.4pts.4D} and~\lemref{dotzero.once} apply unchanged.
The independence of the motion of $v_3$ from $v_1$ permits us
to treat the moving segments $s_0$ and $s_1$ on par with the
fixed segments: the only difference is that their obstacle points
move through $\C_3$ differently.
Therefore a path $\pi_3$ for $v_3(t)$ may be found in $\F_3 \subset \C_3$.
The two paths $\pi_1$ and $\pi_3$, together with the ray $L$ for
$v_2$, constitute a path
for moving the $4$-link chain
$(v_0,v_1,v_2,v_3,v_4)$ through $\C$ while maintaining simplicity.
\end{pf}

\noindent
This finally completes the proof of Theorem~\theoref{line.tracking}.

\subsection{Motion Planning}
\seclab{Canny.2}
We now know a path that avoids self-intersection exists, 
i.e., either the joint $v_1$ or $v_3$ can be straightened.
The next step is to compute such a path algorithmically.
We rely on general motion planning
algorithms, as in Section~\secref{Canny.1}.

Our ``robot'' consists of the four links
$(v_0,v_1,v_2,v_3,v_4)$ moving in the 5-dimensional
configuration space $\C$, Eq.~\eqref{C5}.  
The subspace $\C_0$ that avoids self-intersection between
the four links is some semialgebraic subset of $\C$,
semialgebraic because the constraints on self-intersection 
may be written in Tarski sentences (see, e.g., \cite{m-crag-97}).
The free configuration space $\F$ is composed of the points of $\C_0$
that avoid the obstacles, which is again a semialgebraic set.
Canny's Roadmap algorithm achieves a
time and space complexity of $O(n^5 \log n)$, where $n$ is the number
of obstacles, because in our case, the configuration space
has $k=5$ dimensions.
The algorithm produces a piecewise algebraic
path through $\F$, of $O(n^5)$ pieces. 
Each piece constitutes a constant number of moves,
and so
each joint straightening can be accomplished
in $O(n^5)$ moves.
Repeating the planning and straightening $n$ times
leads to $O(n^6)$ moves in $O(n^6 \log n)$ time.
Because choosing $L$ times requires at most
$O(n^4)$ time by Lemma~\lemref{Choosing.L},
the time complexity is dominated by the path planning,
thereby establishing the bounds claimed in
Theorem~\theoref{closed.4D}.

In the same way that Algorithm~1b improved on Algorithm~1a by avoiding
motion planning, it is likely Algorithm~3 could be improved by an
{\em ad hoc\/} algorithm.

\section{Higher Dimensions}

We have already shown that every simple open chain or tree in 4D can be 
straightened, and every closed chain convexified. 
Our final task is to prove that these results hold for 
higher dimensions, using the results from 4D.

For an open chain, we straighten four links at a time and then repeat the 
procedure until the chain is straight. If the chain or tree
contains fewer than four 
links, then it spans at most a $k$-flat for $k \le 3$, and it can 
be included in $\R^4$. For a closed chain, our algorithm also moves four 
links at a time. Four links determine  at most a $k$-flat $H$ for $k \le 4$
which means that it can be included in a $4$-flat in 
$\R^d$, $d \ge 4$.

We have already shown that these four links, for both all types of chains, 
can be straightened in 4D; therefore, they can be straightened in this 
$4$-flat $H \subset \R^d$. We only have to worry about the pieces of 
the remainder of the chain that 
intersect $H$. But since we are dealing with segments, their 
intersection with $H$ is either a point or a segment. But these are 
the kind of obstructions we have proven that can be avoided in $\R^4$. 
Therefore, 
the straightening of these four links can be completed in $H$, and 
therefore in $\R^d$, while maintaining rigidity and 
simplicity.

The complexity for the algorithms in $\R^d$, $d \ge 4$, is the 
same as for the algorithms in 4D, for all computations are performed
in $4$-flats.
This proves Theorem~\theoref{d.ge.4}.

\vspace{2mm}

\small
\noindent{\bf Acknowledgements.} 
We thank Erik Demaine
and Godfried Toussaint 
for helpful comments, and
Lee Rudolph for help with topology.
We are grateful for the perceptive comments of the referees,
which not only led to
increased clarity throughout, 
but also improved the complexities
of Algorithms~1a and~1b.
\normalsize

\bibliographystyle{alpha}
\bibliography{chains4d}
\end{document}